\documentclass[aps, prd, twocolumn, superscriptaddress, letterpaper, showpacs]{revtex4-1}
\pdfoutput=1
\usepackage{graphicx}
\usepackage{dcolumn}
\usepackage[english]{babel}
\usepackage{lineno}
\usepackage{hyperref}
\usepackage[detect-all]{siunitx}
\usepackage{amssymb,amsmath}
\usepackage[version=4]{mhchem}
\usepackage[caption=false]{subfig}
\usepackage{xcolor}

\makeatletter
\renewcommand{\fnum@figure}{FIG. \thefigure}
\makeatother

\newcommand{\NSixteenEnergy} {6.1\,MeV }
\newcommand{\MuonDeadTime} {20}
\newcommand{\Exposure}{69.2}
\newcommand{\DatasetLivetime} {114.7}
\newcommand{\NumOWLs} {90 }
\newcommand{\LowerEnergyThreshold}{5.0 }
\newcommand{\UpperEnergyThreshold}{15.0 }

\newcommand{\LowBackgroundRate}{$0.25^{+0.09}_{-0.07}$\,events/kt-day}
\newcommand{\HighEnergySolarRate}{$1.03^{+0.13}_{-0.12}$\,events/kt-day}

\newcommand{\SNOFlux}{$\left(5.25\pm0.20\right)$}
\newcommand{\BackgroundRate}{ $10.23\pm0.38$~events/kt-day}
\newcommand{\InteractionRate}{$1.30\pm0.18$~events/kt-day}
\newcommand{\SuperKRate}{$\left(2.345\pm0.039\right)$}
\newcommand{\NueOnlyFlux}{$2.53^{+0.31}_{-0.28}$(stat.)$^{+0.13}_{-0.10}$(syst.)$\times10^6$\,cm$^{-2}$s$^{-1}$}

\newcommand{\beight}[0]{$\ce{^{8}B}$ }
\newcommand{\fluxunits}[1]{#1$\times10^{6}$\,cm$^{-2}$s$^{-1}$}
\newcommand{\figuremacro}[4]{
    \begin{figure}[htbp]
        \centering
        \includegraphics[width=#2\textwidth]{#1}
        \caption[#3]{#4}
        \label{#3}
    \end{figure}
}

\begin{document}
\title{Measurement of the $\ce{^{\textbf{8}}}$B Solar Neutrino Flux in SNO\raisebox{0.35ex}{\small\textbf{+}} with Very Low Backgrounds}

\author{ M.\,Anderson}
\affiliation{\it Queen's University, Department of Physics, Engineering Physics \& Astronomy, Kingston, ON K7L 3N6, Canada}
\author{ S.\,Andringa}
\affiliation{\it Laborat\'{o}rio de Instrumenta\c{c}\~{a}o e  F\'{\i}sica Experimental de Part\'{\i}culas (LIP), Av. Prof. Gama Pinto, 2, 1649-003, Lisboa, Portugal}
\author{ S.\,Asahi}
\affiliation{\it Queen's University, Department of Physics, Engineering Physics \& Astronomy, Kingston, ON K7L 3N6, Canada}
\author{ M.\,Askins}
\affiliation{\it University of California, Davis, 1 Shields Avenue, Davis, CA 95616, USA}
\affiliation{\it University of California, Berkeley, Department of Physics, CA 94720, Berkeley, USA}
\affiliation{\it Lawrence Berkeley National Laboratory, 1 Cyclotron Road, Berkeley, CA 94720-8153, USA}
\author{ D.\,J.\,Auty}
\affiliation{\it University of Alberta, Department of Physics, 4-181 CCIS,  Edmonton, AB T6G 2E1, Canada}

\author{ N.\,Barros}
\affiliation{\it University of Pennsylvania, Department of Physics \& Astronomy, 209 South 33rd Street, Philadelphia, PA 19104-6396, USA}
\author{ D.\,Bartlett}
\affiliation{\it Queen's University, Department of Physics, Engineering Physics \& Astronomy, Kingston, ON K7L 3N6, Canada}
\author{ F.\,Bar\~{a}o}
\affiliation{\it Laborat\'{o}rio de Instrumenta\c{c}\~{a}o e  F\'{\i}sica Experimental de Part\'{\i}culas (LIP), Av. Prof. Gama Pinto, 2, 1649-003, Lisboa, Portugal}
\affiliation{\it Universidade de Lisboa, Instituto Superior T\'{e}cnico (IST), Departamento de F\'{\i}sica, Av. Rovisco Pais, 1049-001 Lisboa, Portugal}
\author{ R.\,Bayes}
\affiliation{\it Laurentian University, Department of Physics, 935 Ramsey Lake Road, Sudbury, ON P3E 2C6, Canada}
\author{ E.\,W.\,Beier}
\affiliation{\it University of Pennsylvania, Department of Physics \& Astronomy, 209 South 33rd Street, Philadelphia, PA 19104-6396, USA}
% ...
\collaboration{The SNO\raisebox{0.5ex}{\tiny\textbf{+}} Collaboration}

\date{\today}

\begin{abstract}
    A measurement of the \beight solar neutrino flux has been made using a
    \Exposure\,kt-day dataset acquired with the SNO\raisebox{0.5ex}{\tiny\textbf{+}} detector during its water
    commissioning phase. At energies above 6\,MeV the dataset is an extremely
    pure sample of solar neutrino elastic scattering events, owing primarily to
    the detector's deep location, allowing
    an accurate measurement with relatively little exposure. In that energy
    region the best fit background rate is \LowBackgroundRate, significantly
    lower than the measured solar neutrino event rate in that energy range,\
    which is \HighEnergySolarRate. Also using data below this threshold, down
    to 5\,MeV, fits of the solar neutrino event direction yielded an observed
    flux of \NueOnlyFlux, assuming no neutrino oscillations. This rate is
    consistent with matter enhanced neutrino oscillations and measurements from
    other experiments.\
\end{abstract}

\maketitle

\section{Introduction}
Neutrinos are produced in the core of the Sun through a variety of nuclear reactions.\
The \beight $\beta^{+}$ decay
(Q$~\approx~$18\,MeV),
 dominates the high-energy portion of the solar neutrino spectrum~\cite{bs05op}.\
Pioneering measurements of the solar neutrino fluxes, including $\ce{^{8}B}$, were made by
 the chlorine and gallium radiochemical experiments~\cite{homestake, sage, gno, gallex},
and the first real-time measurement of solar neutrinos was made by the Kamiokande-II experiment~\cite{kamiokande}.\
The measurement of \beight solar neutrinos by the Sudbury Neutrino Observatory (SNO), along with
measurements of atmospheric and solar neutrinos from Super Kamiokande (Super-K), led to the resolution of the solar neutrino
problem and the initial determination of solar neutrino mixing parameters~\cite{solar_nu_problem,sno_first,sno_second,superk_atmospherics,superk_first_solar}.\
After the first measurements from SNO and Super-K, further \beight solar neutrino
measurements have been made by the liquid scintillator detectors Borexino~\cite{borexino_b8} and
KamLAND~\cite{kamland_solar}.\
These two experiments have also measured solar neutrinos from reactions other than
\beight~\cite{borexino_nature,borexino_final_results,KamlandBe7}.\

Due to the depth and flat overburden at SNOLAB, SNO\raisebox{0.5ex}{\tiny\textbf{+}} has an extremely low rate
of cosmic-ray muons: roughly three per hour. At this rate it is practical to veto
all events for a period of time after each muon (see Sec.~\ref{s:data}) to reduce spallation
backgrounds. As a result, the rate of backgrounds due to cosmogenic
activation and spallation is extremely low.\

This article presents the first solar neutrino results from the SNO\raisebox{0.5ex}{\tiny\textbf{+}}
experiment.\
The low level of backgrounds permits a
measurement of the \beight solar neutrino flux down to 5 MeV with the first 8
months of data. The analysis exercises many tools distinct from those used by
the SNO Collaboration, including new precision
modeling of the detector, energy and vertex reconstruction,
instrumental background rejection, and a well understood level
of intrinsic radioactive contamination in all detector components. These will
be critical to the future sensitivity of SNO\raisebox{0.5ex}{\tiny\textbf{+}} in searches for neutrinoless
double beta decay and measurements of low-energy solar neutrinos~\cite{snop_status_prospects}.

Elastic scattering of electrons by neutrinos,
$\nu_{x}$\,$+$\,$e^{-}$\,$\rightarrow$\,$\nu_{x}$\,$+$\,$e^{-}$ ($x$\,=\,$e$,\,$\mu$,\,$\tau$),
can occur through either a neutral current interaction for neutrinos of all
flavors, or a charged current interaction, for electron neutrinos only.  The
scattered electron's direction is correlated with the direction of the incident
neutrino, so recoil electrons from solar neutrino interactions will typically
produce Cherenkov radiation that is directed away from the Sun. The analysis
presented here exploits this correlation to measure the solar neutrino flux and
spectrum.

\section{Detector}\label{s:detector}
The SNO\raisebox{0.5ex}{\tiny\textbf{+}} detector inherits much of its infrastructure from the SNO
experiment~\cite{sno_detector_paper}.\
The detector is located at a depth of approximately 6000\,m water equivalent
below surface;
it consists of a spherical 6\,m radius acrylic vessel (AV)
suspended within a urylon-lined, barrel-shaped cavity that is 11\,m in radius
and 34\,m tall; the cavity is filled with purified water.\
For the data in this analysis the AV was filled with 0.9\,kt of
``light'' water ($\ce{H_2O}$), as opposed to the heavy water ($\ce{D_{2}O}$) used in SNO.\
Surrounding the AV are 9394 inward-looking 8-inch photomultiplier tubes (PMTs) housed
within a geodesic stainless steel PMT support structure of average radius 8.4\,m.\
Mounted on the outside of the support structure are \NumOWLs outward looking (OWL) PMTs that
serve as a muon veto.\
Each of the SNO\raisebox{0.5ex}{\tiny\textbf{+}} PMTs is surrounded by a
reflective concentrator to increase its effective light collection. A number
of the original SNO PMTs were removed to accommodate a hold-down rope net
that will counteract the buoyant forces on the AV when it is filled with liquid
scintillator~\cite{rope_paper}.\

The PMTs are read out by custom data acquisition (DAQ) electronics that have been largely carried over from
SNO;
parts of the trigger and readout system have been upgraded, allowing a lower
trigger threshold.\
A separate paper discussing in greater detail the SNO\raisebox{0.5ex}{\tiny\textbf{+}} detector is forthcoming.\

\section{Simulation}\label{s:MC}
A Geant4-based~\cite{geant4} Monte Carlo (MC) simulation framework of the
SNO\raisebox{0.5ex}{\tiny\textbf{+}} detector (``RAT'') was used
to determine the expected detector response and selection efficiency for solar neutrino
interactions.\
The \beight neutrino spectrum from Winter \textit{et al.}~\cite{winterspectrum},
and a model of the differential and total cross-sections for electron-neutrino scattering
from Bahcall \textit{et al.}~\cite{escrosssec}, were used to calculate the expected interaction rate.\
The detector simulation models all relevant effects after the initial particle
interaction, including Cherenkov light production, electron scattering processes, photon propagation and
detection, and the DAQ electronics.\
The geometries and material properties in the simulation were determined using
\textit{ex-situ} measurements and \textit{in-situ} calibrations.\
The input parameters for the DAQ simulation were matched to the detector settings
and channel status on a run-by-run basis.\

\section{Reconstruction}\label{s:recon}
For each detected or simulated event, the position, time, direction, and energy
were reconstructed under the assumption that all light produced is Cherenkov
radiation from an electron.\
The direction, time, and position were determined simultaneously through a likelihood fit based
on the pattern and timing of the PMT signals in the event.\
The likelihood was determined using expected distributions of photon timing and
angular spread, which are calculated using MC simulation.\
Only signals originating from well-calibrated channels were used in the fit.\

Energy was determined separately using the position, time, direction, and the
number of PMT signals in a prompt 18\,ns window as inputs;
the prompt time window mitigates the effect of PMT noise and of light
that follows a difficult to model path between creation and detection.\
Using the inputs, the reconstruction algorithm then uses a combination of MC simulation
and analytic calculation to estimate the event energy that is most likely to produce
the observed number of PMT signals.\
The same reconstruction algorithms were used for both simulated and detected
events.\

\begin{figure}[h]
\centering
\dimendef\prevdepth=0
\subfloat[][]{%
\includegraphics[width=0.5\textwidth]{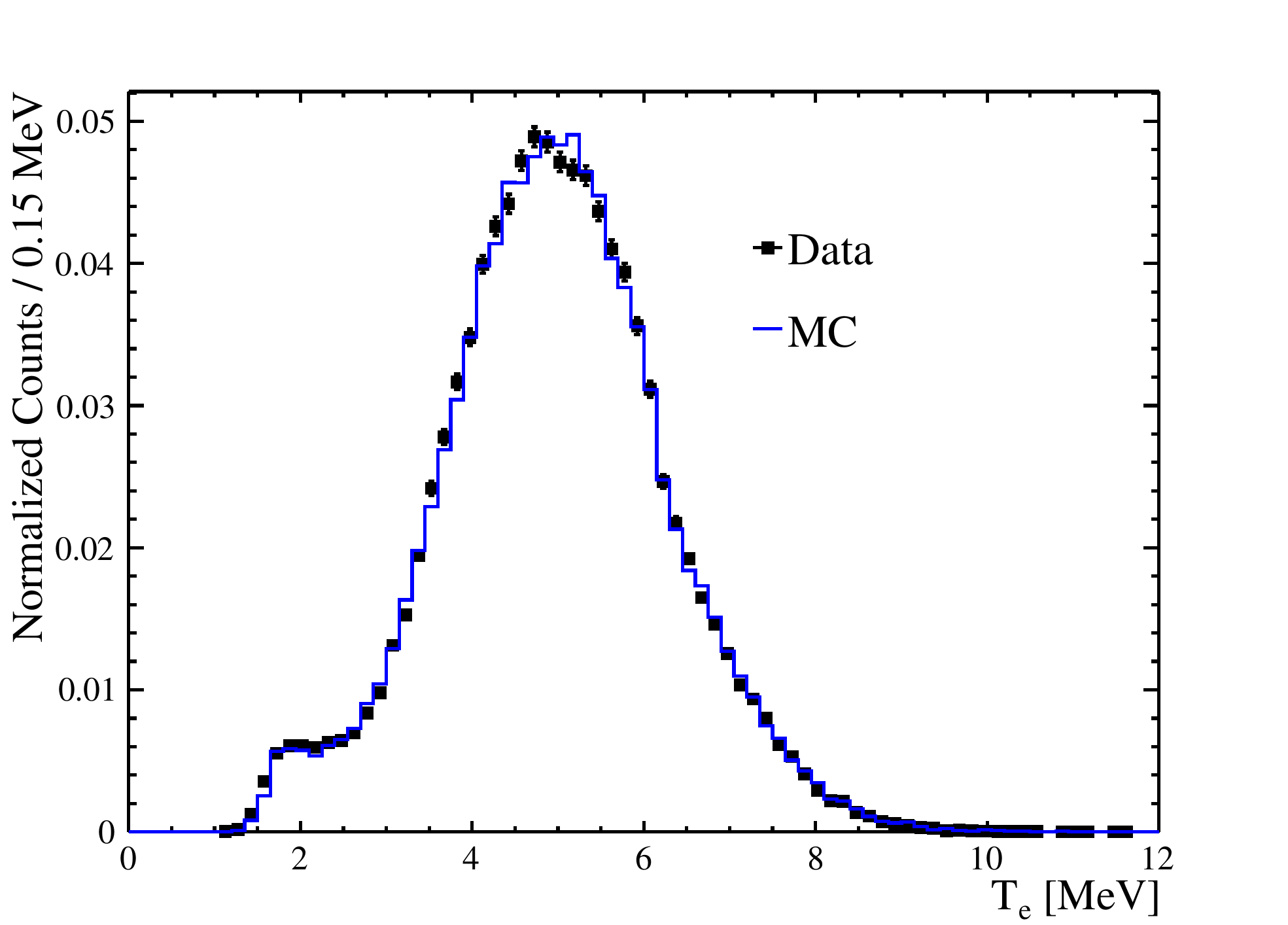}%
\label{fig:n16Energy}%
}
\hfill
\subfloat[][]{%
\includegraphics[width=0.5\textwidth]{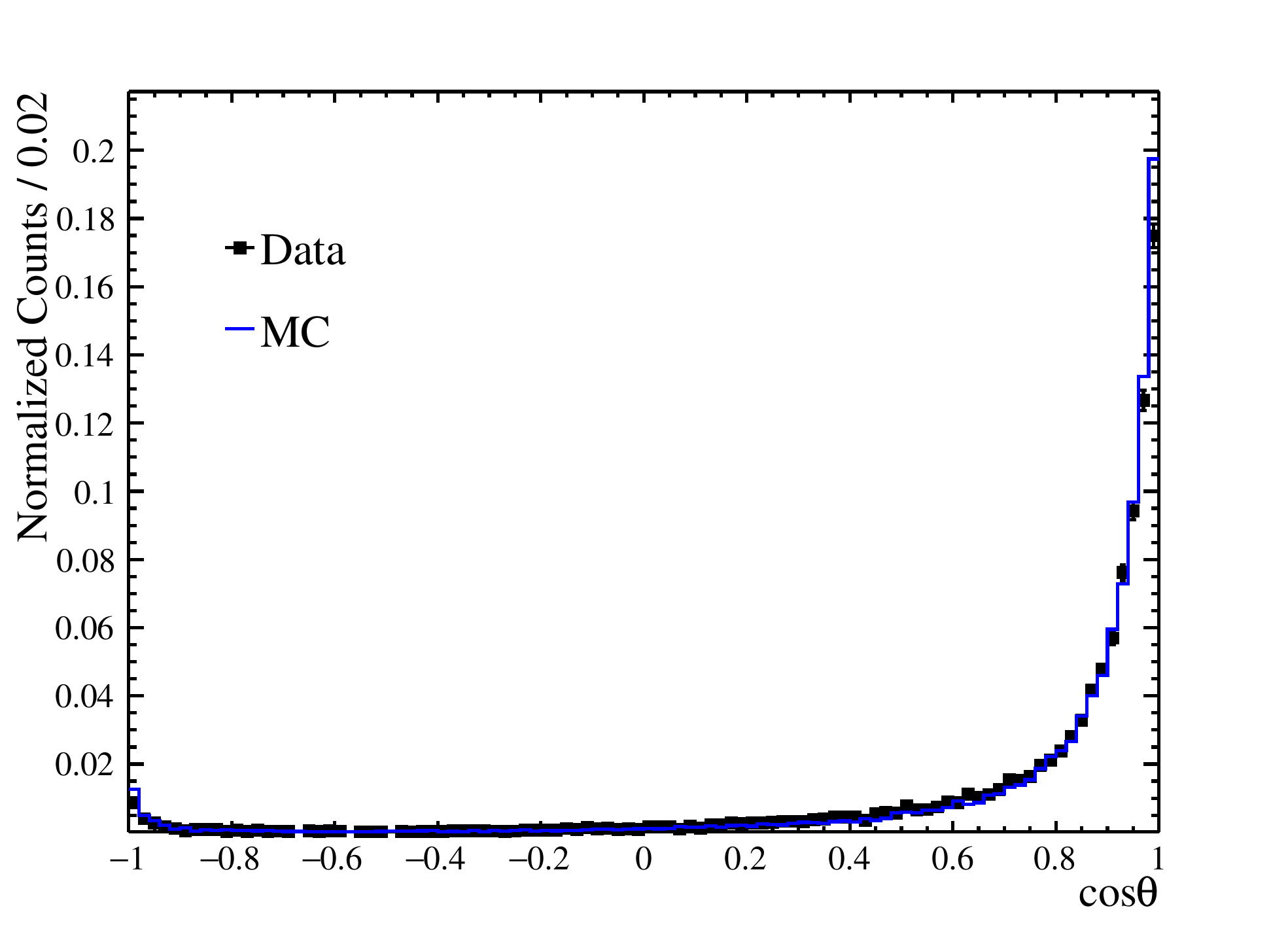}%
\label{fig:n16_direction}%
}
\caption{Energy\,(a) and direction\,(b) reconstruction comparison between
    Monte Carlo simulation and data for $\ce{^{16}N}$ events. Here $T_{e}$ is the
    electron kinetic energy, and $\theta$ is
    the angle between the reconstructed direction of an event and the direction vector pointing
    from the source position to the event's reconstructed position.}
\label{fig:n16}%
\end{figure}

\section{Calibration}\label{s:calib}
Calibration data were taken with a deployed $\ce{^{16}N}$ source~\cite{snoN16},
which primarily produces a tagged \NSixteenEnergy  $\gamma$-ray.\
These data are used for calibrating detector components and evaluating systematic
uncertainties of reconstructed quantities.\

The source position was controlled using a system of ropes to perform a
3-dimensional scan of the space inside the AV, and a 1-dimensional vertical scan
in the region between the AV and the PMTs. For the purpose of evaluating systematics, the
distributions of events in position, direction, and energy were fitted with
response functions. The parameters from fits performed on data and MC
simulation were compared to assign systematic uncertainties on each
reconstructed quantity. Figure~\ref{fig:n16} shows comparisons between simulation and data
for the reconstructed energy and direction of $\ce{^{16}N}$ events.\

The 6.1\,MeV $\ce{^{16}N}$ $\gamma$-ray typically Compton scatters in the detector
to produce one or more electrons that reconstruct to energies peaking near 5\,MeV.\
In addition to Compton scattering, energy deposition in the source
container also produces a substantial tail at lower energies. This tail fades out
below about 1.7\,MeV due to the detector trigger thresholds (see Fig.~\ref{fig:n16Energy}). The
energy resolution is composed of several effects including Compton scattering,
detector resolution, and photon statistics; the latter being dominant. In the
fit of the energy response function, the detector resolution was modeled as
Gaussian and convolved with an $\ce{^{16}N}$ energy spectrum determined from MC
simulation to account for the other two components. The resulting
fractional uncertainty on the resolution within the fiducial volume and
at kinetic energy $T_{e}$ is $0.018\sqrt{T_{e}/\text{MeV}}$;
the fractional energy scale uncertainty is $2.0\%$.\
Similarly, for the position fit, the response function includes a convolution
with the angular distribution of photon production to account for the
non-negligible mean free path of the $\ce{^{16}N}$ $\gamma$-ray. The photon
production distribution was also determined from MC simulation.\
More information about the $\ce{^{16}N}$ source analysis is
available in Ref.~\cite{snop_nd}, in which
other water phase physics results from SNO\raisebox{0.5ex}{\tiny\textbf{+}} are presented.

\section{Dataset}\label{s:data}
Data for this analysis were gathered from May through December, 2017.\
Calibrations and detector maintenance were also performed during this period.\
Data taking periods were split into runs; the typical run length was between 30 and 60 minutes.\
Each run was checked against a number of criteria to ensure its quality.\
This included checks on the spatial uniformity of PMT signals, trigger rate, laboratory activity, and
detector stability.\

Within each run, muons and interactions from atmospheric neutrinos were tagged
using the number of OWL PMT signals in an event and the number of events that follow closely in time.\
After each muon or atmospheric event, a \MuonDeadTime~second deadtime
was introduced to reduce backgrounds from cosmogenically produced isotopes, such as $\ce{^{16}N}$.\
Additional adjustments to the overall livetime were made to account for
removal of time-correlated instrumental backgrounds.\
The resulting dataset contains 120 days of data and a corresponding livetime of
\DatasetLivetime~days, or \Exposure\,kt-days exposure with the fiducial volume
cut described in Sec.~\ref{s:analysis}.\

\section{Analysis}\label{s:analysis}
For each event, a suite of low-level cuts were applied to reject events originating
from instrumental effects, and to ensure that the events had energies
high enough to lie in a region of well-understood and near-perfect trigger
efficiency.\
The trigger efficiency cut requires the number of PMT signals in a 100\,ns
coincidence window to be above a certain threshold. During the first 60\% of
dataset livetime, the threshold for this cut was $23$, while the trigger
threshold itself was $15$ in-time signals. For the remaining section of data, the
trigger threshold was lowered to $7$ in-time PMT signals and the corresponding
trigger efficiency cut was $10$.\

For events passing the low level cuts, it was further required that the vertex reconstruction
fits successfully converged.\
Unsuccessful fits can occur if an event takes place in an optically complicated
region of the detector, e.g., near the cylindrical chimney at the top of the AV.\
These regions often distort the light distribution from an event such that its vertex cannot be
reliably determined.\

A fiducial volume cut was then introduced requiring that each event reconstruct
within 5.3\,m of the detector center,
reducing backgrounds from events originating on or outside the AV. A
more restrictive cut on position was used for the beginning of data taking to
minimize the impact of an increased rate of external backgrounds in the upper
half of the detector.\
For that data, events observed in the upper half of the detector were required
to be within 4.2\,m of the center. The more restrictive cut was applied for
$13\%$ of the dataset livetime.\

After vertex reconstruction, additional cuts were placed on the timing and isotropy of PMT signals
in each event. These cuts removed residual contamination from instrumental
backgrounds (which have neither the prompt timing nor angular distribution of Cherenkov light), as well as events with
poorly fit vertices.\
The timing cut required at least 55\% of the PMT signals occur within
a time-of-flight corrected prompt time window of width 7.5\,ns.\
Isotropy was parameterised by $\beta_{14}$, a value determined by the
first and fourth Legendre polynomials of the angular distribution of PMT signals within
an event~\cite{sno_leta}.\
Events were required to have $\beta_{14}$ is between $-0.12$ and $0.95$.\

A final cut was placed on the reconstructed kinetic energy of each event, selecting only events
within the energy region \LowerEnergyThreshold to \UpperEnergyThreshold\,MeV,
removing most of the backgrounds from radioactivity and atmospheric neutrino
interactions; the only solar neutrinos with a significant flux in this energy
region are \beight neutrinos.\
The fiducial volume and energy cuts select $21.4\%$ of simulated solar
$\nu_e$ events that interact within the AV; events were simulated according to
the \beight energy spectrum.\
The efficiencies of the other cuts on events that are within the energy region
and fiducial volume are given in Table~\ref{table:mceff}.\
Table~\ref{table:dataeff} shows the effect of each cut on the dataset.\

\begin{table}
    \setlength{\tabcolsep}{4pt}
\begin{center}
\begin{tabular}{l r}
\hline
\hline
    Selection & Passing MC Fraction\\
\hline
    Total (after energy \& position cuts) & 1.0 \\
    Low-level cuts & 0.988 \\
    Trigger Efficiency  & 0.988 \\
    Hit Timing  & 0.988 \\
    Isotropy & 0.986 \\
\hline
\hline
\end{tabular}
    \caption{Efficiency for each cut on MC simulated solar $\nu_{e}$ events that are within
    the fiducial volume and the energy region.}
\label{table:mceff}
\end{center}
\end{table}

\begin{table}
    \setlength{\tabcolsep}{4pt}
\begin{center}
\begin{tabular}{l r}
\hline
\hline
    Selection & Passing Triggers\\
\hline
    Total & 12 447 734 554\\
    Low-level cuts & 4 547 357 090\\
    Trigger Efficiency & 126 207 227\\
    Fit Valid & 31 491 305\\
    Fiducial Volume  & 6 958 079\\
    Hit Timing  & 2 752 332\\
    Isotropy & 2 496 747\\
    Energy  & 820\\
\hline
\hline
\end{tabular}
\caption{Dataset reduction for each applied cut. The second column is the number
    of triggered events from the detector that pass each cut.}
\label{table:dataeff}
\end{center}
\end{table}

Since the direction of the recoil electron in a solar neutrino scattering event
is correlated with the position of the Sun, the rate of solar neutrino events in
the dataset was extracted by fitting the distribution of events in
$\cos\theta_{\text{sun}}$, where
$\theta_\text{{sun}}$ is the angle between an event's reconstructed direction
and the vector pointing directly away from the Sun at the time of the event.\
The rate of radioactive backgrounds present in the dataset can be determined as one of the
parameters in the fit,\
so no \textit{a priori} knowledge of the background rate was required.\

Events with reconstructed kinetic energy, $T_{e}$, between $5.0$ and
$10.0$\,MeV were distributed among five uniformly wide bins, and a single bin from
$10.0$ to $15.0$\,MeV.\
In each energy bin, a maximum likelihood fit was performed on the distribution of events in
$\cos\theta_\text{{sun}}$ to determine the rate of solar neutrino events and the rate
of background events as a function of energy.\
The expected distribution for solar neutrino events in $\cos\theta_\text{{sun}}$ was calculated from MC simulation.\
The PDF for background events was taken to be uniform in $\cos\theta_\text{{sun}}$.\
The best fit flux over all energies was found by maximizing the product
of the likelihoods from the fit in each energy bin.\
The resulting likelihood function is given by
\begin{align}
    \mathcal{L}(S&, \bold{B}, \delta_{\theta} | \bold{n}, \mu_{\theta}, \sigma_{\theta}) = \nonumber\\
    &\mathcal{N}(\delta_{\theta}, \mu_{\theta}, \sigma_\theta)
    \prod_{j=0}^{N_E}\prod^{N_{\theta}}_{i=0} \text{Pois}\left(n_{ij},\, B_{j} + S\,p_{ij}(\delta_{\theta})\right)\text{.}
    \label{eq:ll}
\end{align}
 The number of energy bins and angular bins are represented by
 $N_{E}$ and $N_{\theta}$ respectively.\
$S$ is the solar neutrino interaction rate and is the parameter of interest for this analysis,
$B_{j}$ is the background rate in each energy bin.\
$\mathcal{N}$ represents a normalized Gaussian distribution.\
The $\delta_{\theta}$ parameter represents an adjustment to the angular
resolution; $\mu_{\theta}$ and $\sigma_{\theta}$ are respectively the best fit and the
constraint on $\delta_{\theta}$ from the $\ce{^{16}N}$ source analysis.\
The number of observed counts in the $i^{th}$ angular bin and $j^{th}$ energy
bin is given by $n_{ij}$, and $p_{ij}(\delta_{\theta})$ is the corresponding
predicted solar probability density for a given angular resolution parameter.\
$\text{Pois}\left(k, \lambda \right)$ is the value of the Poisson distribution
at the value $k$ for a rate parameter $\lambda$.\

Systematic uncertainties were propagated by varying the reconstructed quantities
for each simulated event.\
A fit was then performed with each modified solar PDF to determine the effect
the systematic uncertainty has on the final result.\
Because this analysis relies heavily on direction reconstruction, the angular
resolution ($\delta_{\theta}$) was treated as a nuisance parameter in the fit for the solar flux.\
Details about the systematic uncertainties can be found in Ref.~\cite{snop_nd}.

\figuremacro{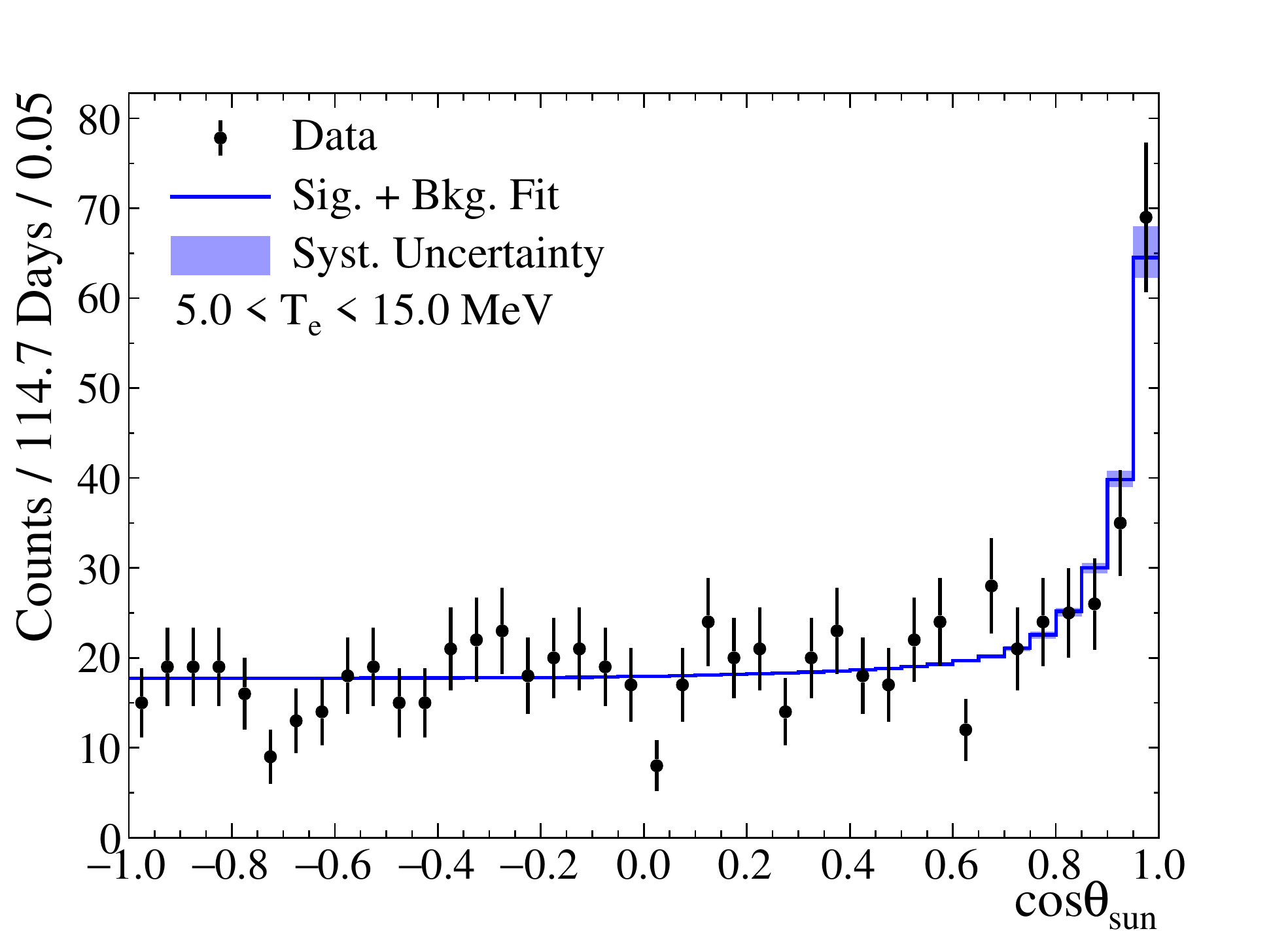}{0.5}{fig:costheta}{Distribution of event direction
                                              with respect to solar direction.\
                                              The systematic error bar includes angular correlated and
                                              uncorrelated errors.}

\section{Results}\label{s:results}
Figure~\ref{fig:costheta} shows the distribution of events in $\cos\theta_\text{{sun}}$
for events over the entire energy range of $5$ to $15$\,MeV and the fit to that distribution.\
The fit gives a solar event rate of \InteractionRate\
 and background rate of \BackgroundRate.\
Performing a similar fit in each individual energy bin yielded a best fit solar flux
as a function of energy.\
The fits were combined, in accordance with Eq.~\ref{eq:ll}, yielding an overall best fit flux of
\begin{equation*}
    \Phi_{ES}= 2.53^{+0.31}_{-0.28}\text{(stat.)}^{+0.13}_{-0.10}\text{(syst.)}\times10^6\,\text{cm}^{-2}\text{s}^{-1}\text{.}
\end{equation*}
This value assumes the neutrino flux consists purely of electron flavor neutrinos.\
The result agrees with the elastic scattering flux published by Super-K,
$\Phi_{ES}$=\fluxunits{\SuperKRate}~\cite{superk4}, combining statistical and
systematic errors.\

\figuremacro{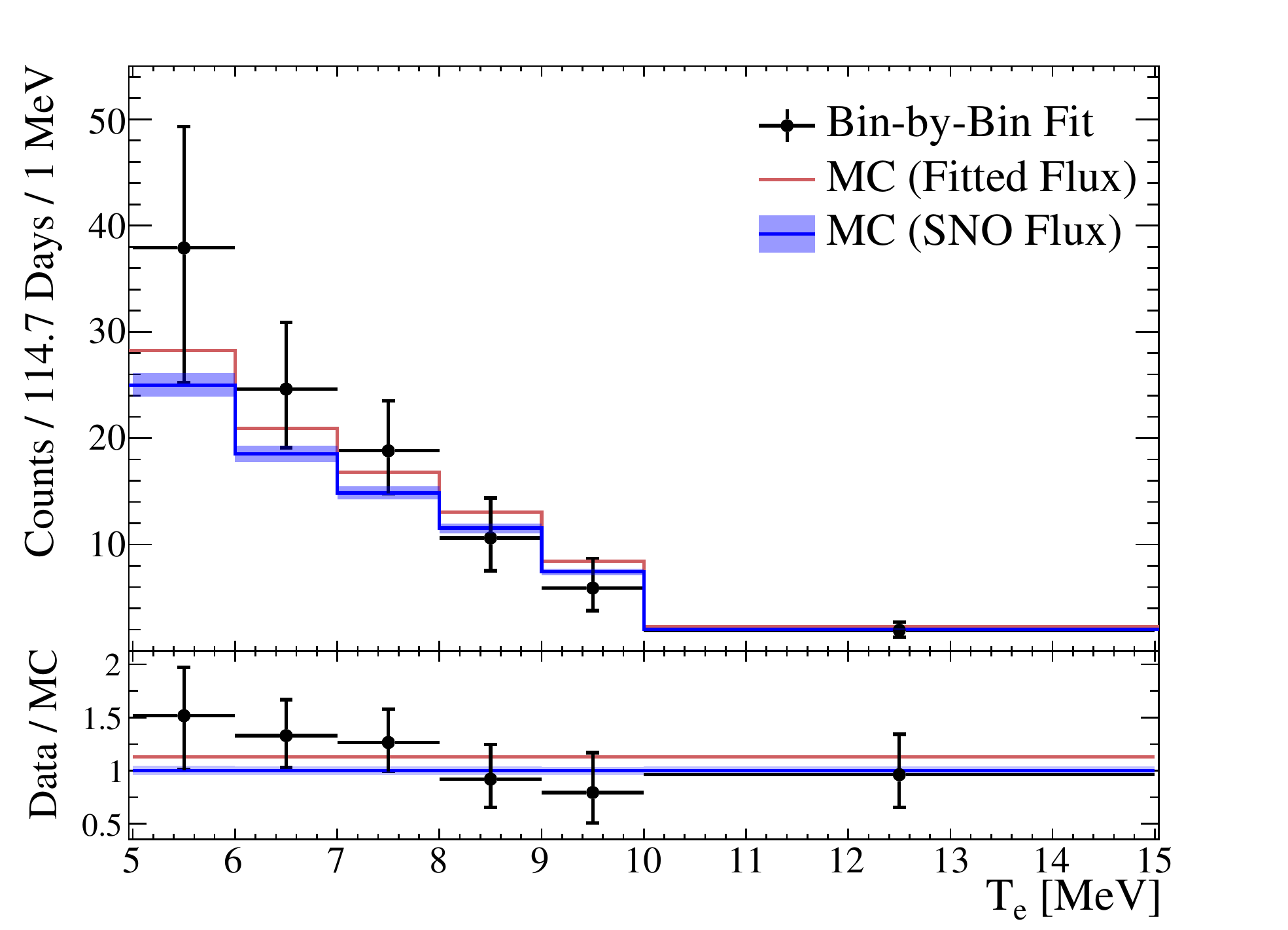}{0.5}{spectrum}{
    (Top) The extracted solar neutrino elastic scattering event rate as a
    function of reconstructed electron kinetic energy $T_{e}$.\
    (Bottom) The same, as a fraction of the expected rate.\
    The red and blue lines show the MC simulation predicted spectrum normalized
    to the best fit flux and the SNO flux measurement~\cite{sno_combined}, respectively.\
    The uncertainty on the SNO result includes reported uncertainty combined with
    mixing parameter uncertainties.\
    The black points are the results of the fits to the $\cos\theta_\text{{sun}}$
    distribution in each energy bin,
    with error bars indicating the combined statistical and systematic
    uncertainty, including energy-correlated uncertainty.\
    A horizontal dash is placed on each error bar indicating the statistics
    only uncertainty; for all points the statistical error is dominant and the systematic
    error bar is not visible above the dash.}

\begin{table}
\begin{center}
\begin{tabular}{l c}
\hline
\hline
Systematic & Effect \\
\hline
Energy Scale & 3.9\% \\
Fiducial Volume & 2.8\% \\
Angular Resolution & 1.7\% \\
Mixing Parameters & 1.4\% \\
Energy Resolution & 0.4\% \\
\hline
Total & 5.0\%\\
\hline
\hline
\end{tabular}
\caption{Effect of each systematic uncertainty on the extracted solar neutrino
         flux. Systematic uncertainties with negligible effects
         are not shown. For asymmetric uncertainties, the larger is shown.}
\label{table:systematics}
\end{center}
\end{table}

Including the effects of solar neutrino oscillations, using the neutrino mixing
parameters given in Ref.~\cite{pdg2016} and the solar production and electron
density distributions given in Ref.~\cite{bs05op} gave a best fit solar flux
of
\begin{equation*}
    \Phi_{\ce{^{8}B}}= 5.95^{+0.75}_{-0.71}\text{(stat.)}^{+0.28}_{-0.30}\text{(syst.)}\times10^{6}\text{cm}^{-2}\text{s}^{-1}\text{.}
\end{equation*}
This result is consistent with the \beight flux as measured by the SNO experiment,
$\Phi_{\ce{^{8}B}}$=\fluxunits{\SNOFlux}~\cite{sno_combined}, combining statistical
and systematic uncertainties.\
Figure~\ref{spectrum} shows the best fit solar neutrino \beight event rate in each
energy bin along with the predicted energy spectrum scaled to the best fit
flux, and scaled to the flux measured by SNO. Each statistical error bar on the
measured rate is affected by both the solar neutrino and background rates in that
energy bin.\
Table~\ref{table:systematics} details how each systematic uncertainty affects this result.\

\begin{figure}[htbp]
    \centering
\includegraphics[width=0.5\textwidth]{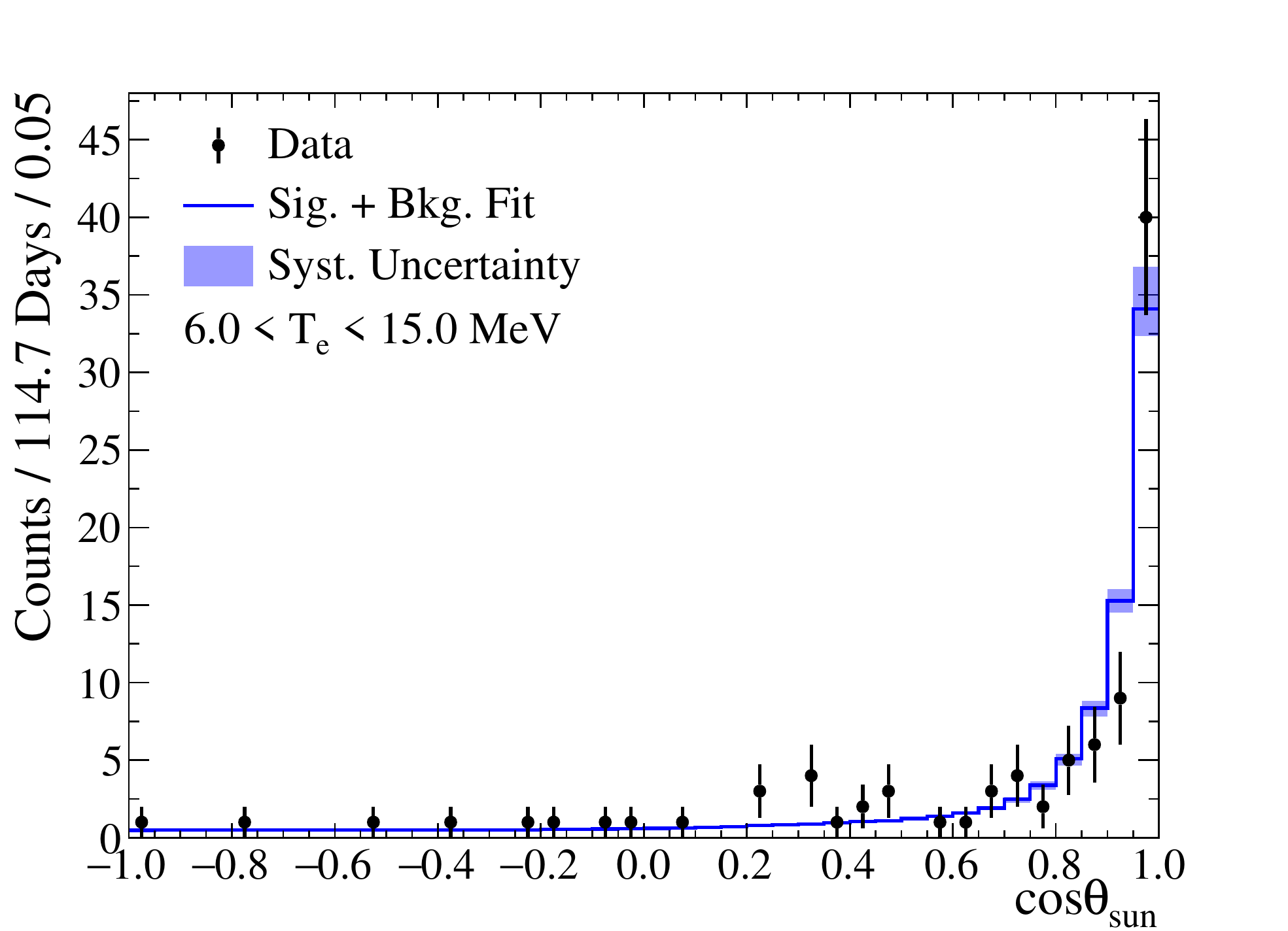}%
\caption{Distribution of event directions with respect to solar direction for
    events with energy in \numrange[range-phrase=--]{6.0}{15.0}\,MeV.}
\label{fig:cos_theta_six}
\end{figure}

The upper five energy bins, \numrange[range-phrase=--]{6.0}{15.0}\,MeV, were an
extremely low background region for this analysis.\
There was very little background contamination from
cosmogenically produced isotopes due primarily to depth of the detector.\
The comparatively high rate of backgrounds in the \numrange[range-phrase=--]{5.0}{6.0}\,MeV bin
comes primarily from decays of radioactive isotopes, such as radon, within the detector.\
Figure~\ref{fig:cos_theta_six} shows the distribution in $\cos\theta_\text{{sun}}$ of events at
energies above 6~MeV, illustrating the low background rate.\
In that energy region the best fit background rate was \LowBackgroundRate, much
lower than the measured solar rate in that energy range, \HighEnergySolarRate.\
For the region above 6~MeV, this is the lowest
background elastic scattering measurement of solar neutrinos in a water
Cherenkov detector.\

\section{Conclusion}\label{s:conc}
Described here is the first measurement of the \beight solar neutrino flux as observed
by the SNO\raisebox{0.5ex}{\tiny\textbf{+}} detector in its water phase using \DatasetLivetime\,days of data.\
Our results are consistent with measurements from other experiments,
and serve to provide continued monitoring of reactions within the core of the Sun.\

The low rate of backgrounds above 6\,MeV, in conjunction with the measured systematic uncertainties,
allows an accurate measurement of the solar neutrino flux despite the limited size of the dataset.\
The low background rates at high energies come primarily from a low rate of cosmic-ray
muons within the detector volume,
and allows the measurement of other physical phenomena, including invisible
nucleon decay~\cite{snop_nd} and potentially the local reactor anti-neutrino flux in the SNO\raisebox{0.5ex}{\tiny\textbf{+}} water
phase.\
The presence of radon backgrounds from the internal water limits this analysis at lower energies.\
In SNO\raisebox{0.5ex}{\tiny\textbf{+}}'s scintillator and tellurium loaded phases
the internal radioactive backgrounds will
be significantly reduced, allowing further measurements of solar neutrinos at lower energies.\

\section*{Acknowledgements}

Capital construction funds for the SNO\raisebox{0.5ex}{\tiny\textbf{+}} experiment were provided by the Canada
Foundation for Innovation (CFI) and matching partners.\
This research was supported by:
{\bf Canada:}
Natural Sciences and Engineering Research Council,
the Canadian Institute for Advanced Research (CIFAR),
Queen's University at Kingston,
Ontario Ministry of Research, Innovation and Science,
 Alberta Science and Research Investments Program,
National Research Council,
 Federal Economic Development Initiative for Northern Ontario,
Northern Ontario Heritage Fund Corporation,
Ontario Early Researcher Awards;
{\bf US:}
Department of Energy Office of Nuclear Physics,
National Science Foundation,
 the University of California, Berkeley,
Department of Energy National Nuclear Security Administration through the
Nuclear Science and Security Consortium;
{\bf UK:}
Science and Technology Facilities Council (STFC),
the European Union's Seventh Framework Programme under the European Research
Council (ERC) grant agreement,
the Marie Curie grant agreement;
{\bf Portugal:}
Funda\c{c}\~{a}o para a Ci\^{e}ncia e a Tecnologia (FCT-Portugal);
{\bf Germany:}
the Deutsche Forschungsgemeinschaft;
{\bf Mexico:}
DGAPA-UNAM and Consejo Nacional de Ciencia y Tecnolog\'{i}a.\

We thank the SNO\raisebox{0.5ex}{\tiny\textbf{+}} technical staff for their strong contributions. We would
like to thank SNOLAB and its staff for support through underground space,
logistical and technical services. SNOLAB operations are supported by the
CFI and the Province of Ontario Ministry of
Research and Innovation, with underground access provided by Vale at the
Creighton mine site.\

This research was enabled in part by support provided by WestGRID
(www.westgrid.ca) and Compute Canada (www.computecanada.ca) in particular
computer systems and support from the University of Alberta (www.ualberta.ca)
and from Simon Fraser University (www.sfu.ca) and by the GridPP Collaboration,
in particular computer systems and support from Rutherford Appleton Laboratory~\cite{gridpp, gridpp2}.\
 Additional high-performance computing was provided
through the ``Illume'' cluster funded by the CFI
and Alberta Economic Development and Trade (EDT) and operated by
ComputeCanada and the Savio computational cluster resource provided by the
Berkeley Research Computing program at the University of California, Berkeley
(supported by the UC Berkeley Chancellor, Vice Chancellor for Research, and
Chief Information Officer). Additional long-term storage was provided by the
Fermilab Scientific Computing Division. Fermilab is managed by Fermi Research
Alliance, LLC (FRA) under Contract with the U.S. Department of Energy, Office
of Science, Office of High Energy Physics.\
\bibliography{references}
\end{document}